\documentclass[aps,preprint,pdftex]{revtex4}
\usepackage{amssymb}
\usepackage{amsmath}
\usepackage{graphicx}
\usepackage{bm}
\usepackage{color}
\usepackage{subfigure}

\begin{document}
\title{
Formation of edge pressure pedestal and reversed magnetic shear due to toroidal rotation in a tokamak equilibrium
}
\author{Haolong Li}
\affiliation{CAS Key Laboratory of Geospace Environment and Department of Engineering and Applied Physics \\ University of Science and Technology of China \\ Hefei, Anhui 230026, China}
\author{Ping Zhu}
\email[E-mail:]{zhup@hust.edu.cn}
\affiliation{International Joint Research Laboratory of Magnetic Confinement Fusion and Plasma Physics, State Key Laboratory of Advanced Electromagnetic Engineering and Technology, School of Electrical and Electronic Engineering, Huazhong University of Science and Technology, Wuhan, Hubei 430074, China \\ \\ Department of Engineering Physics \\  University of Wisconsin-Madison\\ Madison, Wisconsin 53706, USA}
\date{\today}

\begin{abstract}

Toroidal rotation is well known to play significant roles in the edge transport and L-H transition dynamics of tokamaks. Our recent calculation finds that a sufficiently strong localized toroidal rotation can directly bring out the formation of edge pressure pedestal with reversed magnetic shear that is reminiscent of an H-mode plasma, purely through the effects of toroidal rotation on the tokamak MHD equilibrium itself. In particular, the enhanced edge toroidal rotation enables a substantial peaking of the parallel current profile near edge in higher $\beta$ regimes, which leads to the flattening or reversal of the local $q$ (safety factor) profile. Here the formation of pressure pedestal along with the reversed magnetic shear region is shown to be the natural outcome of the MHD tokamak equilibrium in a self-consistent response to the presence of a localized toroidal rotation typically observed in H-mode or QH-mode.

\end{abstract}

\maketitle
\section{Introduction}

High confinement mode (H-mode) \cite{prl_wangner_1982} and quiescent H-mode (QH-mode) in tokamaks are often associated with the strong local toroidal flow in the edge region \cite{Liang_2020,ppcf_burrell_2002,Oyama_2005,Sakamoto_2004,Suttrop_2005,Suttrop_2003}, which can be generated intrinsically or driven by the auxiliary heating method, for example, Neutral Beam Injection(NBI) \cite{Rice_2016}. Both experiments and theory suggest that the edge sheared toroidal flow can contribute to the formation of edge transport barrier \cite{Kobayashi_2020} and H-mode \cite{McKee_2009} through its suppression of microinstability turbulence \cite{Liang_2020}. The stabilizing effects of sheared toroidal rotation on the edge localized mode (ELM) \cite{Aiba2009,Furukawa2005,Waelbroeck1991,Cooper1988,Aiba2010, Aiba2011,xi2012,xia2013} may eventually lead to the emergence of QH-mode and edge harmonic oscillation (EHO) \cite{Chatthong_2010,Snyder_2007}. However, the direct role of toroidal rotation in the establishment of MHD equilibrium of H- or QH-modes is less well-known.

There is a rich history of research on the self-consistent equilibrium with plasma flow. The axisymmetric MHD equations in the presence of macroscopic poloidal and toroidal flows are derived in various forms by a number of authors \cite{flow_infls2,flow_infls4,NongXiang_ref1,flow_infls1,analysis_author1}. Besides, several codes has been developed to compute equilibria with plasma flow, such as CLIO \citep{code_clio} DIVA \cite{code_diva1}, FINESSE \cite{code_finesse}, FLOW \cite{NongXiang_ref1} and a version of FLOW that includes contributions from nonthermal populations \cite{FLOW_version1}. The modified Grad-Shafranov equation due to toroidal flow is first given concisely by Hameiri \cite{flow_infls2}. Maschke and Perrin later are able to obtain the analytic solution of the modified Grad-Shafranov equation after some simplifying assumptions \cite{flow_infls1}. The dominant flow effect previously found on tokamak equilibrium is the outward shifts of magnetic surfaces and plasma profiles. In particular, Guazzotto \emph{et al.} have investigated numerically the effect of toroidal and poloidal flows on the equilibrium of tokamak plasma using the equilibrium code FLOW \cite{NongXiang_ref1}, and concluded that a density pedestal around the edge is generated when the poloidal velocity exceeds the local sound speed \cite{guazzotto_prl}.


In this work, we study the direct effects of edge sheared toroidal rotation on the MHD equilibrium of an L-mode. It is found that the rigid toroidal flow shifts plasma in the direction of major radius. Inclusion of an edge localized toroidal rotation enables the formation of a pedestal structure in the pressure and number density profiles as a consequence of the self-consistent tokamak MHD equilibrium. For higher $\beta$ tokamak, an edge region with zero or reversed magnetic shear can appear as the localized toroidal rotation increases in its peak magnitude. It turns out the H-mode edge pedestal along with a non-monotonic $q$ profile is a natural and self-consistent MHD equilibrium response to the presence of strongly edge localized toroidal flow.

The paper is organized as follows: The Grad-Shafranov equation with toroidal rotation is discussed in Section \ref{sec:nimeq_analy}. Pedestal structure formation due to strong sheared toroidal flow is reported in Section \ref{sec:pedestal_formation}, which is followed by an account of the related $q$ profile flattening and magnetic shear reversal. in Section \ref{sec:q_flatted}.  In the end, conclusion and discussion are presented in Section \ref{sec:summary}.

\section{Grad-Shafranov equation and solutions in presence of toroidal rotation}
\label{sec:nimeq_analy}

In presence of toroidal flow $\vec{u} = R^{2} \Omega (\psi) \nabla \phi$, the Grad-Shafranov equation becomes \cite{Aiba_codes_ref,furukawa_flow,nimeq_my_2020}

\begin{eqnarray}
\Delta^{*} \psi = - \mu_{0} R^{2} \frac{\partial P}{\partial \psi} - F \frac{\mathrm{d}F}{\mathrm{d} \psi}
\label{eq:nimeq_gs}
\\
P(\psi ,R)=P_{0}(\psi) \exp\left[ \frac{m_{i}R_{0}^{2} \Omega^{2}}{2T} \left( \frac{R^{2}}{R_{0}^{2}} -1 \right) \right]
\label{eq:nimeq_p}
\\
\rho(\psi ,R)=\rho_{0}(\psi) \exp\left[ \frac{m_{i}R_{0}^{2} \Omega^{2}}{2T} \left( \frac{R^{2}}{R_{0}^{2}} -1 \right) \right]
\label{eq:nimeq_rho}
\end{eqnarray}

where the Grad-Shafranov operator is defined as 
\begin{eqnarray}
\Delta^{*} \equiv R \frac{\partial}{\partial R} R^{-1} \frac{\partial}{\partial R} + \frac{\partial^{2}}{\partial Z ^{2}}
\label{eq:nimeq_analy3}
\end{eqnarray}
where $\rho$ is the mass density, $\vec{u}$ is the toroidal rotation velocity, $P$ is the plasma pressure, $\vec{J}$ is the current density, $\vec{B}$ is the magnetic field, $\mu_{0}$ is the permeability of space, $m_{i}$ is the mass of ion, $T$ is the plasma temperature, $R$ is the major radius, $\phi$ is the toroidal angle, $\Omega(\psi)$ is the frequency of toroidal rotation, which is a flux function judged from the curl of Ohm's law. $\psi$ is the physical poloidal flux divided by a factor of $2\pi$. Meanwhile, $P_{0}(\psi)$ is considered as the pressure profile without toroidal rotation.

Substituting Eq. (\ref{eq:nimeq_p}) into Eq. (\ref{eq:nimeq_gs}) yields
\begin{eqnarray}
\nonumber
&& \Delta^{*} \psi= - F \frac{\mathrm{d}F}{\mathrm{d}\psi} - \mu_{0} R^{2} \exp\left[ \frac{m_{i}R_{0}^{2} \Omega^{2}}{2T} \left( \frac{R^{2}}{R_{0}^{2}} -1 \right) \right] \\
&& \left[\frac{\mathrm{d}P_{0}}{\mathrm{d}\psi} + P_{0} \frac{m_{i} R_{0}^2 \Omega}{T} \left(\frac{R^{2}}{R_{0}^{2}}-1\right) \frac{\mathrm{d}\Omega}{\mathrm{d}\psi} - P_{0} \frac{m_{i} R_{0}^{2} \Omega^{2}}{2T^{2}} \left(\frac{R^{2}}{R_{0}^{2}}-1\right) \frac{\mathrm{d}T}{\mathrm{d}\psi} \right]
\label{eq:nimeq_gs_all_term}
\end{eqnarray}

For tokamak plasma the thermal conduction along magnetic field lines is fast compared to the heat transport perpendicular to a magnetic surface. Thus, plasma temperature can be considered as a flux function, namely $T=T(\psi)$. Due to $P(\psi,R)=n(\psi,R)T(\psi)$, there will be a shift of number density profile in the direction of major radius, similar to pressure profile, as shown in both analytic and numerical solutions \cite{nimeq_my_2020}. Physically, this shift arises from the centrifugal force induced by toroidal flow, namely $\rho (\vec{u} \cdot \nabla) \vec{u} = - \rho R \Omega^{2} \hat{e}_{R}$, where $\hat{e}_{R}$ denotes the direction along the major radius $R$. Thus, toroidal flow modifies equilibrium through both the pressure profile and the  poloidal magnetic flux, as governed by Eq. (\ref{eq:nimeq_p}) and Eq. (\ref{eq:nimeq_rho}).

\begin{figure}[ht]
\subfigure[]{
  \includegraphics[width=0.44\textwidth]{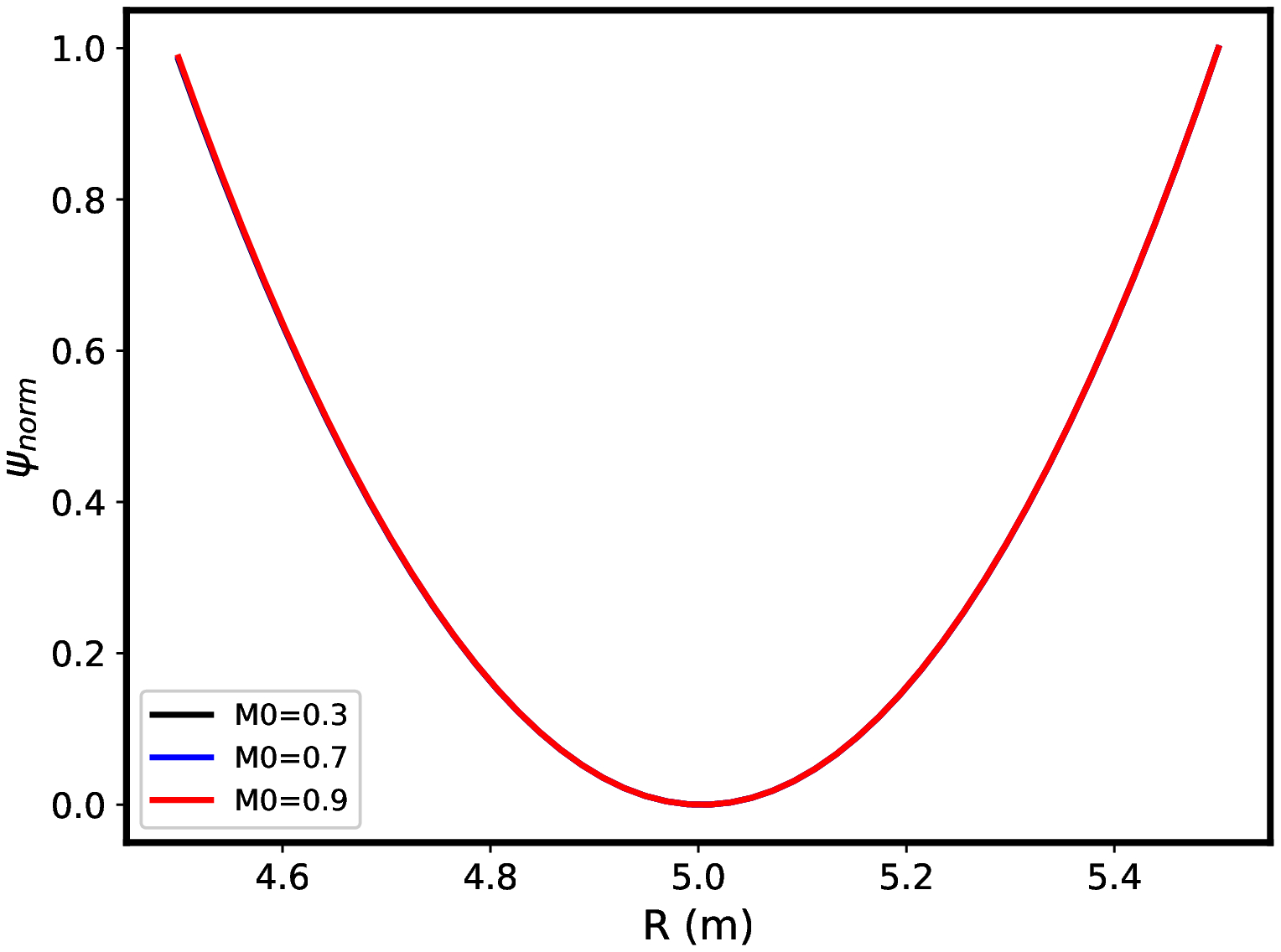}
  }
  \subfigure[]{
  \includegraphics[width=0.44\textwidth]{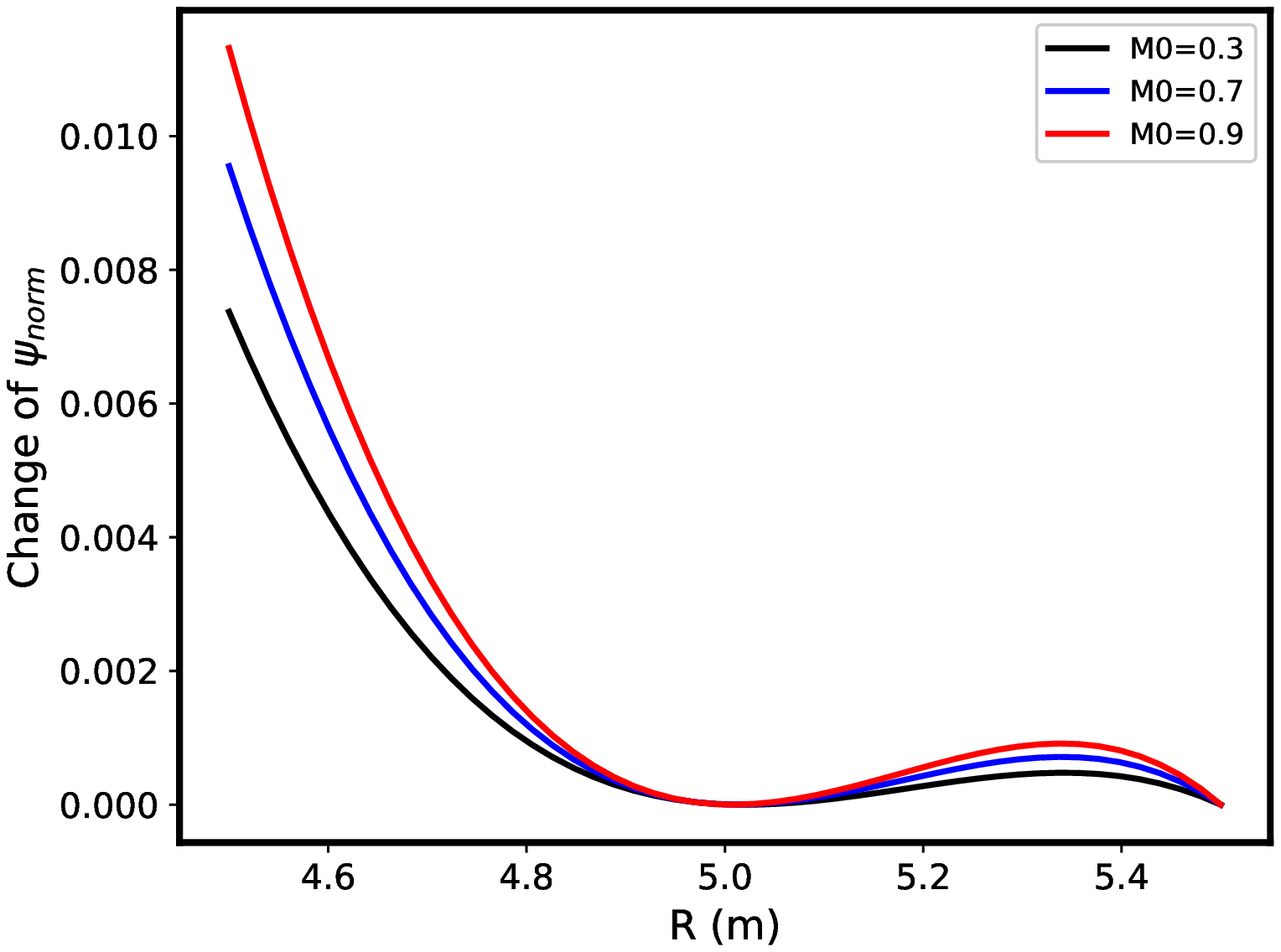}
}

\subfigure[]{
  \includegraphics[width=0.44\textwidth]{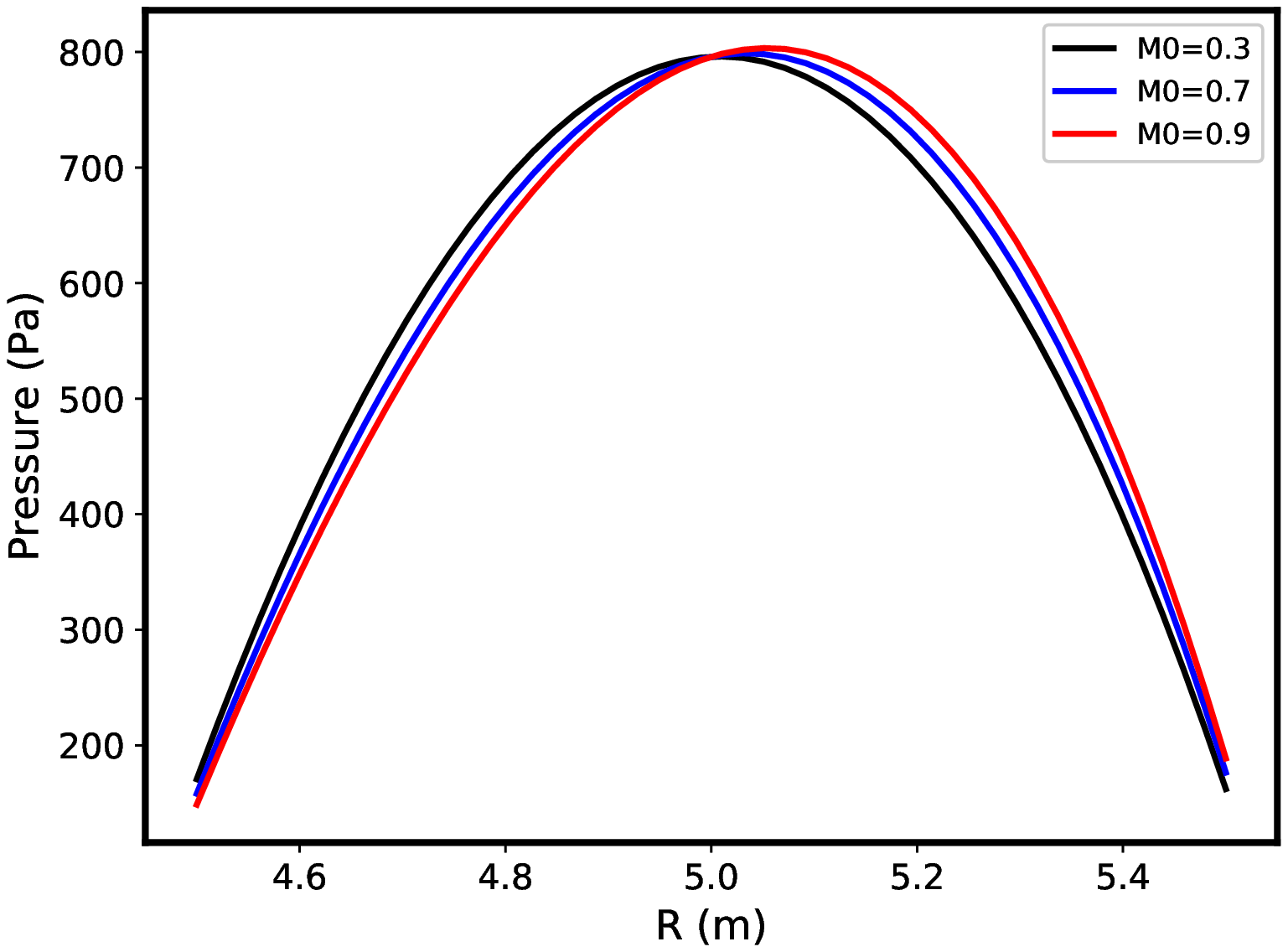}
  \label{fig:scan_M0_analy_myself_prof_prof}
}\subfigure[]{
  \includegraphics[width=0.44\textwidth]{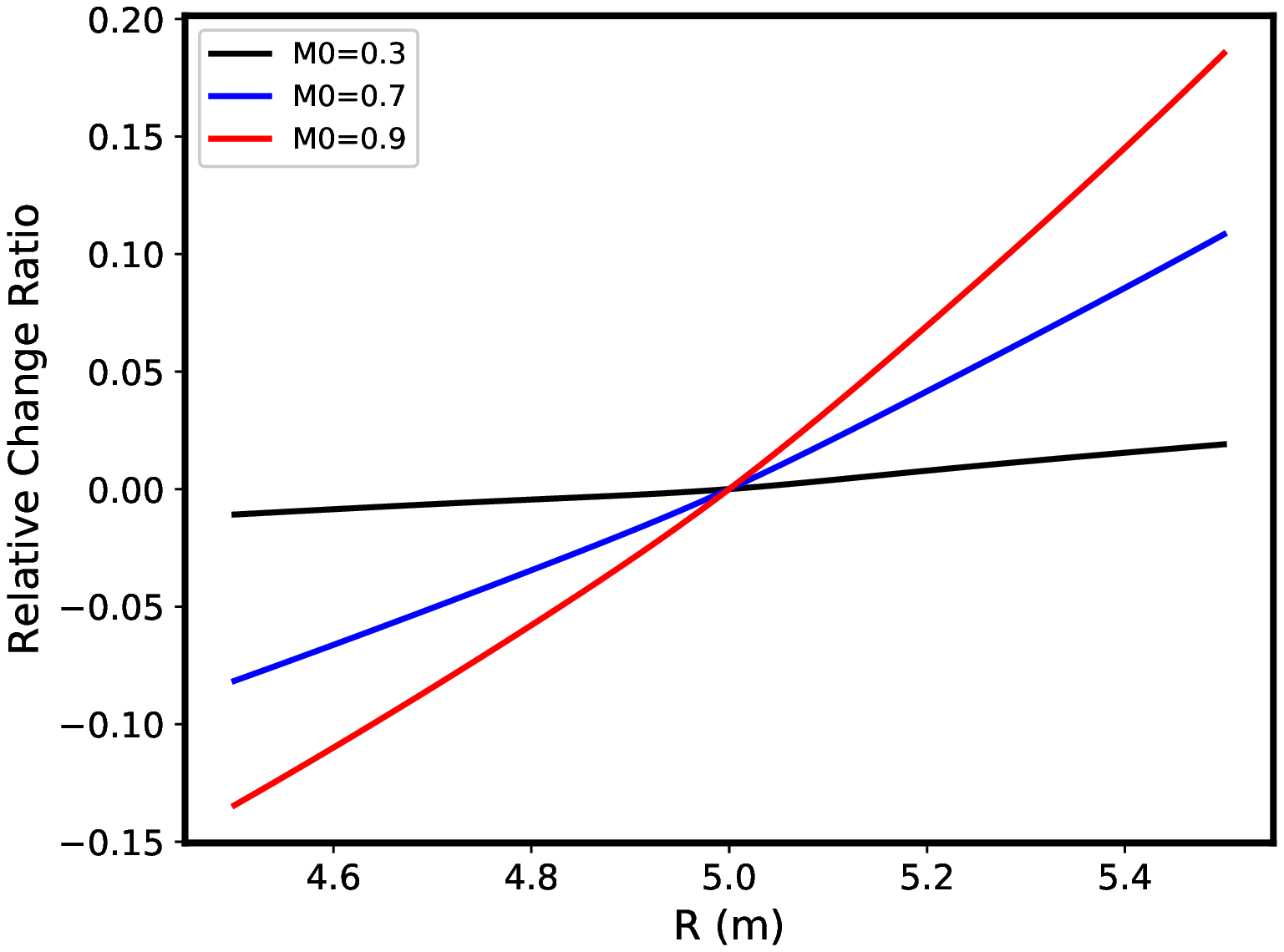}
  \label{fig:scan_M0_analy_myself_prof_ratio}
}

\subfigure[]{
  \includegraphics[width=0.44\textwidth]{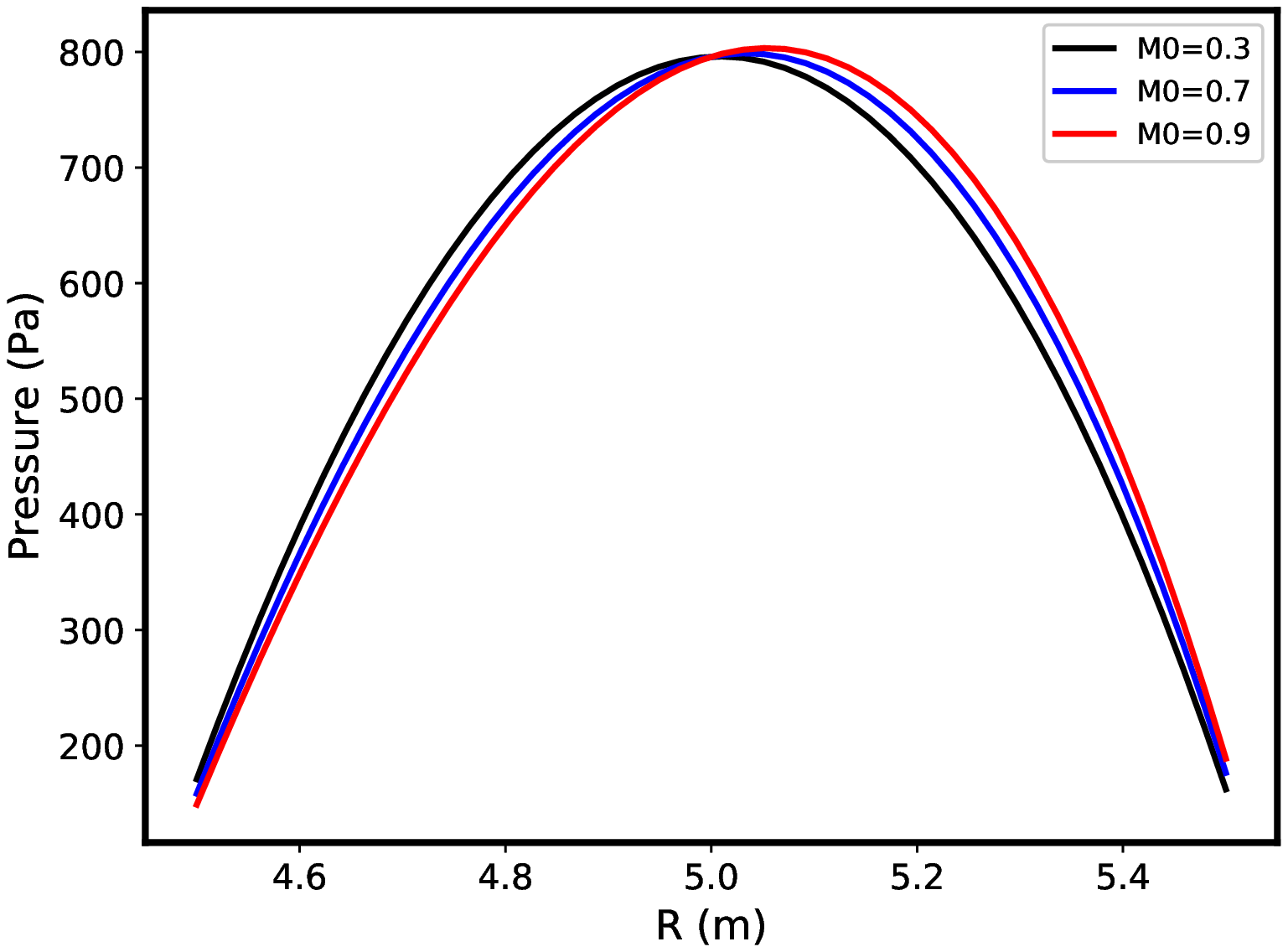}
  \label{fig:scan_M0_analy_myself_all_prof}
}\subfigure[]{
  \includegraphics[width=0.44\textwidth]{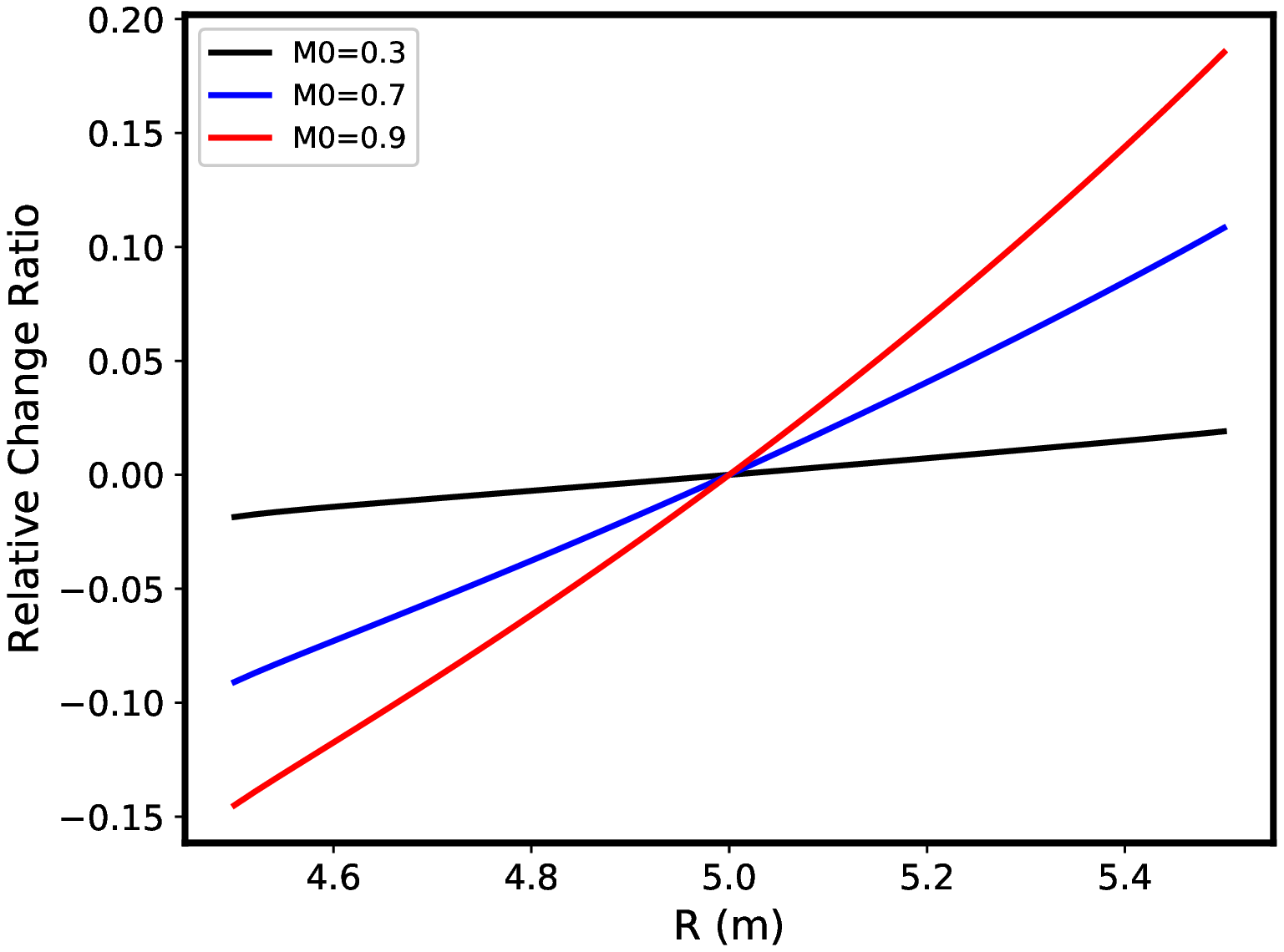}
  \label{fig:scan_M0_analy_myself_all_ratio}
}

\caption{(a) The normalized poloidal flux function $\psi$ from the analytical solution of Solov'ev equilibrium in presence of rigid toroidal flow, and (b) its change from the static equilibrium solution for different Mach numbers. The corresponding pressure and its relative change from the static equilibrium as functions of the major radius (c-d) ignoring and (e-f) including self-consistently the change in $\psi$ due to toroidal rotation. The computational domain is a rectangular cross section with $4.5m<R<5.5m$ and $-0.5m<Z<0.5m$, and the equilibrium parameters $p_{0}=1.0 \times 10^{-3}$, $p_{1}=-8.0 \times 10^{-4}$, $F_{0}=5.0$.}

\label{fig:scan_M0_analy_myself}
\end{figure}

Toroidal flow alone directly introduces a shift of pressure profile in the $R$ direction. As can be seen from Eq. (\ref{eq:nimeq_p}), because of the toroidal flow, the pressure is reduced in the strong field side where $R<R_{0}$, i.e. $P(\psi, R) <P_{0}(\psi)$. Similarly, in the weak field side, $R>R_{0}$, the pressure increases, i.e. $P(\psi, R) > P_{0}(\psi)$. As a consequence, the overall peak of pressure profile shifts towards the weak field side due to the toroidal flow. This can be demonstrated using the Solov'ev equilibrium for example. For the static Solov'ev equilibrium, the pressure profile is $\mu_{0} P_{0}=p_{0}+p_{1} \psi$ and the $F$ profile satisfies $F \frac{\rm{d} F}{\rm{d} \psi}=F_{0}$. In presence of a rigid toroidal flow, the analytical solution for Solov'ev equilibrium becomes \cite{nimeq_my_2020}
\begin{eqnarray}
\label{eq:analy_solov_rigid}
\psi=\psi_{h}+\psi_{p}=c_{1}+c_{2}R^{2}+c_{3}(R^{4}-4R^{2}Z^{2})+c_{4}[R^{2}\ln(R)-Z^{2}] 
\\
\nonumber
-p_{1} \left( \frac{R_{0}^{2}}{2M_{0}^{2}} \right)^{2} \left\lbrace \exp \left[M_{0}^{2}\left(\frac{R^{2}}{R_{0}^{2}}-1\right)\right] - M_{0}^{2} \left( \frac{R^{2}}{R_{0}^{2}} - 1 \right) - 1 \right\rbrace-\frac{F_{0}}{2}Z^{2}
\end{eqnarray}
where $M_{0}= \frac{m_{i}R_{0}^{2}\Omega_{0}^{2}}{2T_{0}}$ denotes the Mach number at $R=R_{0}$. As the rigid toroidal flow increases in magnitude, the pressure profile based on the analytical solution in Eq. (\ref{eq:analy_solov_rigid}) indeed skews outward (Fig. \ref{fig:scan_M0_analy_myself}).

Figs. \ref{fig:scan_M0_analy_myself}(a)-\ref{fig:scan_M0_analy_myself}(b) also show that the change of poloidal magnetic flux induced by toroidal flow is quite small, which is consistent with Eq. (\ref{eq:analy_solov_rigid}). Comparing the last two terms of Eq. (\ref{eq:analy_solov_rigid}), we have
\begin{eqnarray}
\frac{-p_{1} \left( \frac{R_{0}^{2}}{2M_{0}^{2}} \right)^{2} \left\lbrace \exp \left[M_{0}^{2}\left(\frac{R^{2}}{R_{0}^{2}}-1\right)\right] - M_{0}^{2} \left( \frac{R^{2}}{R_{0}^{2}} - 1 \right) - 1 \right\rbrace}{-\frac{F_{0}}{2}Z^{2}} \sim \frac{p_{1} L_{R}^{4}}{\frac{F_{0}}{2} L_{Z}^{2}} \sim \frac{p_{1}}{\frac{B_{\phi}^{2}}{2}} \sim \beta
\label{eq:terms_compare}
\end{eqnarray}
where $L_{R}$ and $L_{Z}$ denote the characteristic length of equilibrium in the horizontal and the vertical direction, respectively, with the assumption that $L_{R}=L_{Z}$. For low $\beta$ equilibrium, the change of $\psi$ due to rigid toroidal flow may be negligible. More generally, comparing the two terms on the right hand side of Eq. (\ref{eq:nimeq_gs}) leads
\begin{eqnarray}
\frac{- \mu_{0} R^{2} \frac{\partial P}{\partial \psi}}{-F\frac{\mathrm{d}F}{\mathrm{d} \psi}} \sim \frac{P}{\frac{B_{\phi}^{2}}{2\mu_{0}}} \sim \beta
\label{eq:terms_compare_gs}
\end{eqnarray}
Since the toroidal flow only appears explicitly in the term $- \mu_{0} R^{2} \frac{\partial P}{\partial \psi}$ of the extended Grad-Shafranov equation, the change  in the equilibrium flux function $\psi$ induced by toroidal flow is expected to be small for a low-$\beta$ equilibrium.

\section{Formation of edge pressure pedestal due to localized toroidal flow}
\label{sec:pedestal_formation}

\begin{figure}[ht]

\subfigure[]{
  \includegraphics[width=0.35\textwidth, height=0.23\textheight]{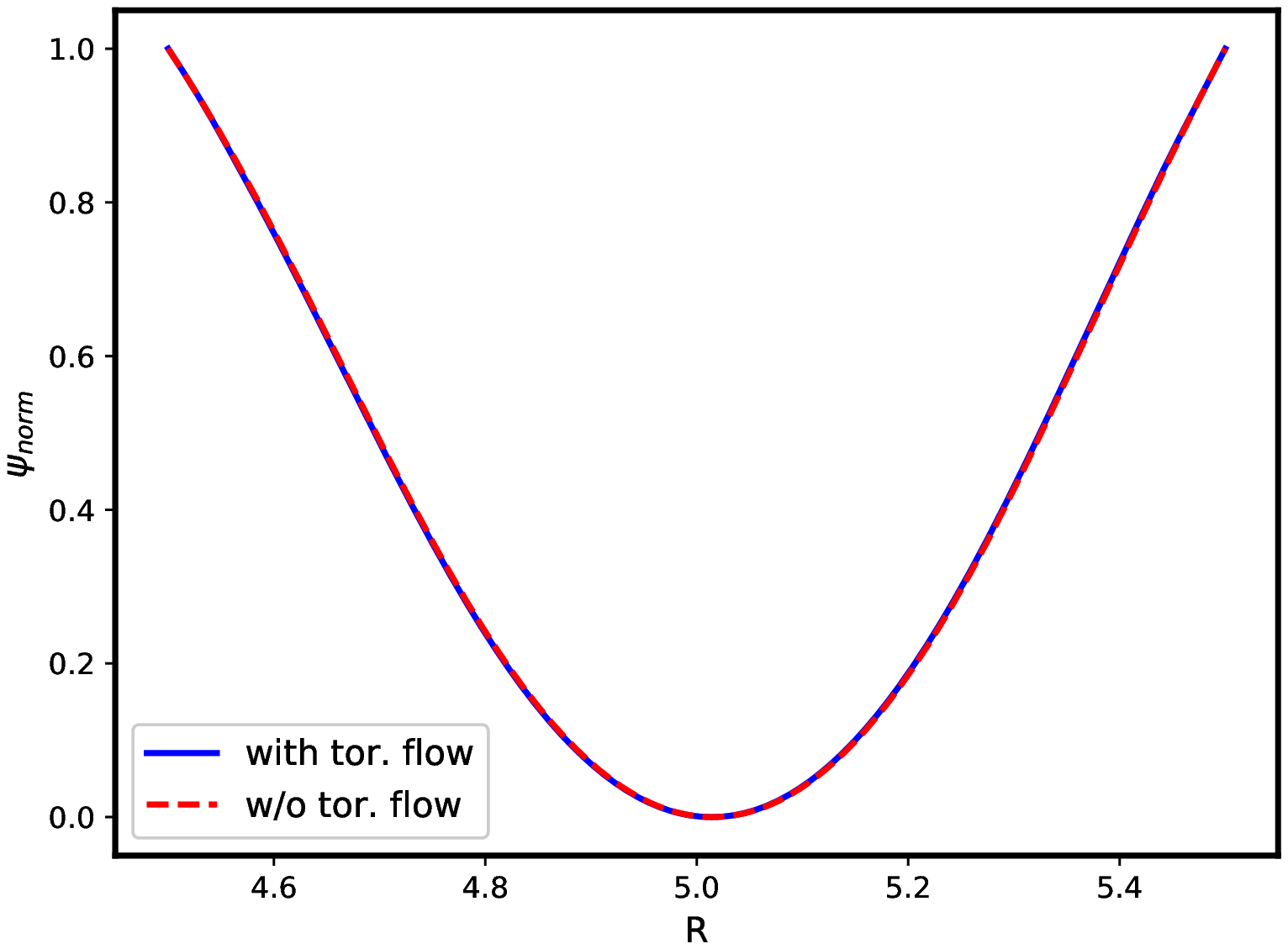}
  \label{fig:separate_psi_prof_effects_psi_1}
}\subfigure[]{
  \includegraphics[width=0.35\textwidth, height=0.23\textheight]{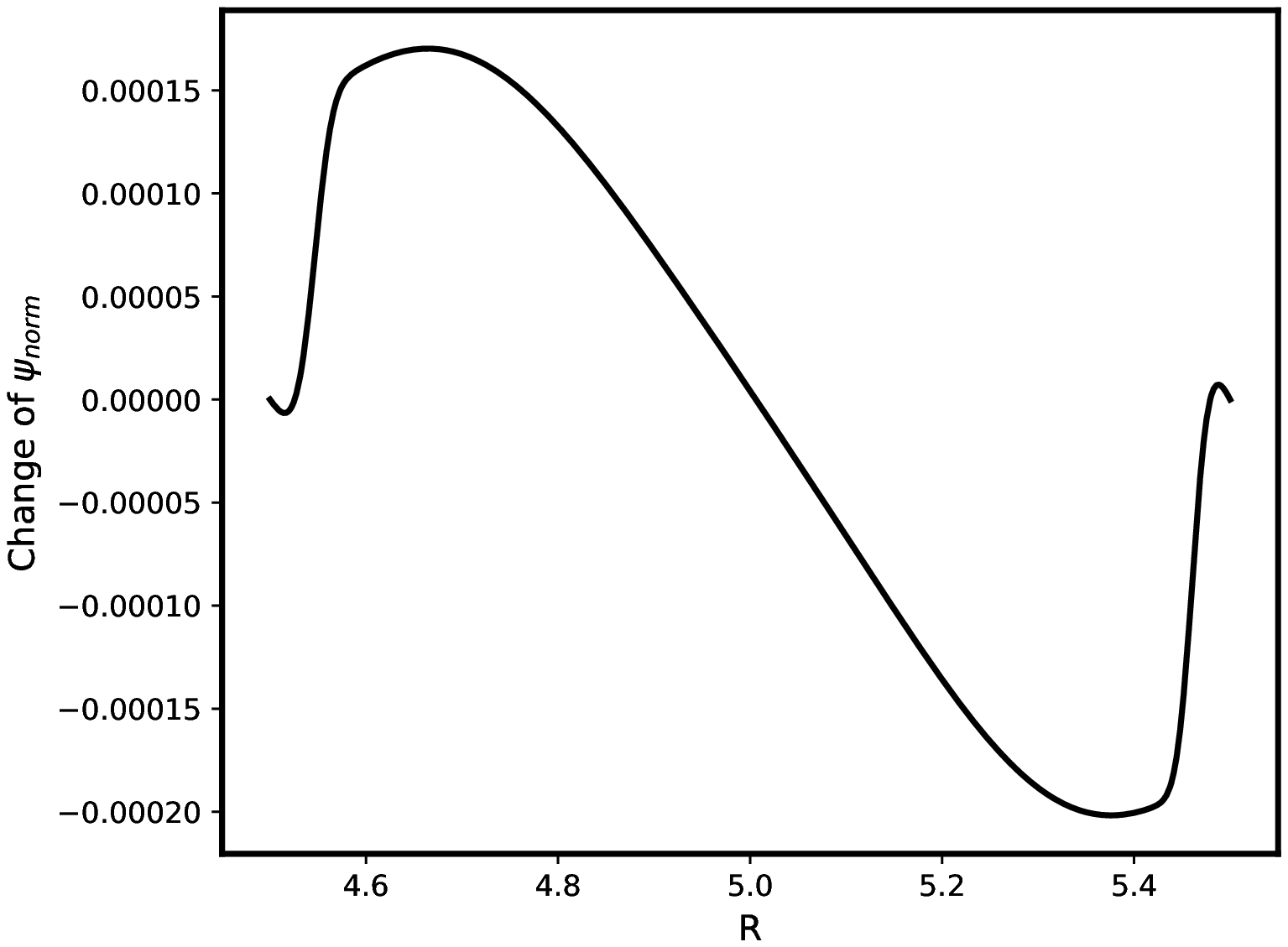}
  \label{fig:separate_psi_prof_effects_psi_2}
}

\subfigure[]{
  \includegraphics[width=0.35\textwidth, height=0.23\textheight]{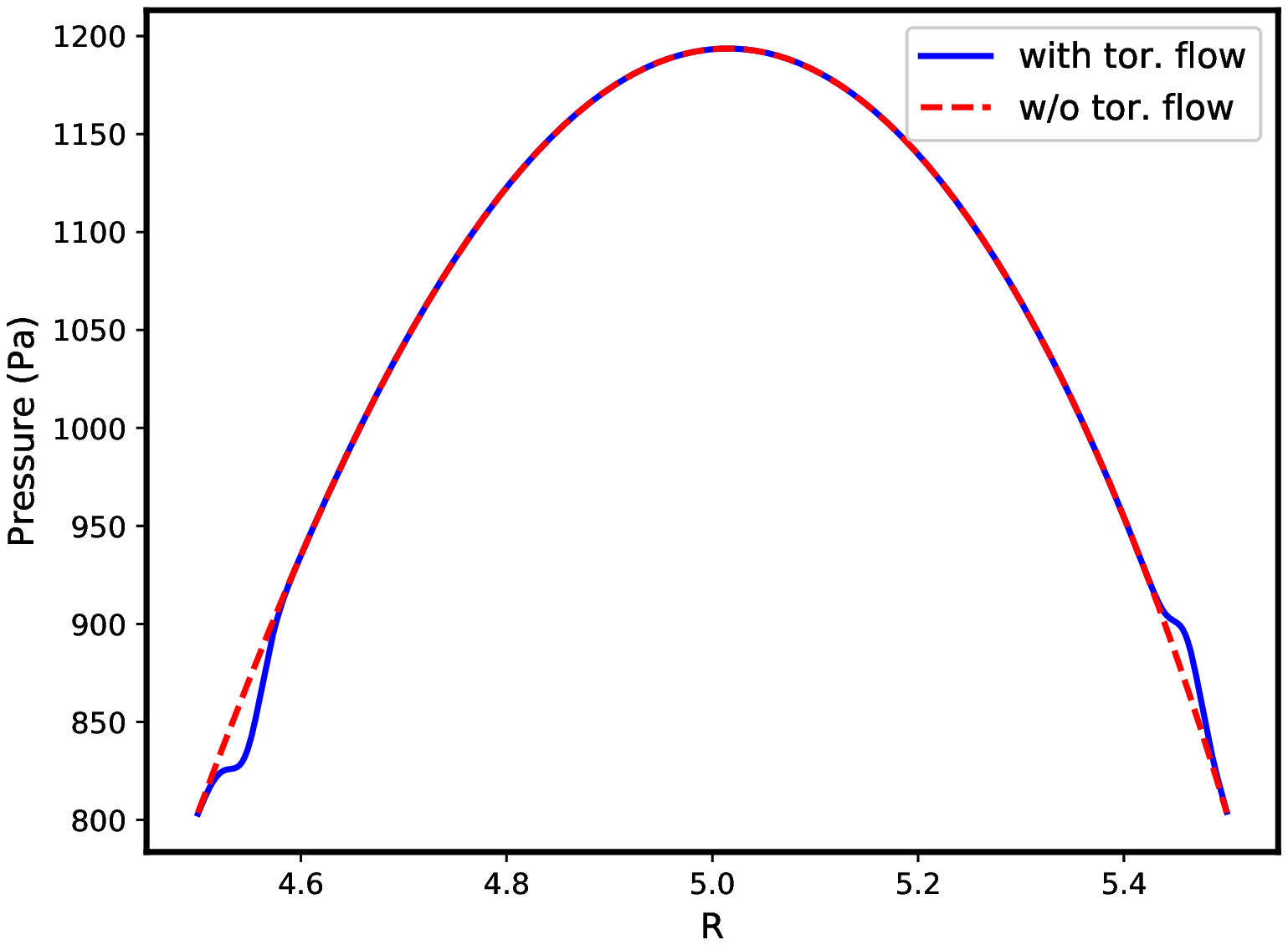}
  \label{fig:separate_psi_prof_effects_prof_1}
}\subfigure[]{
  \includegraphics[width=0.35\textwidth, height=0.23\textheight]{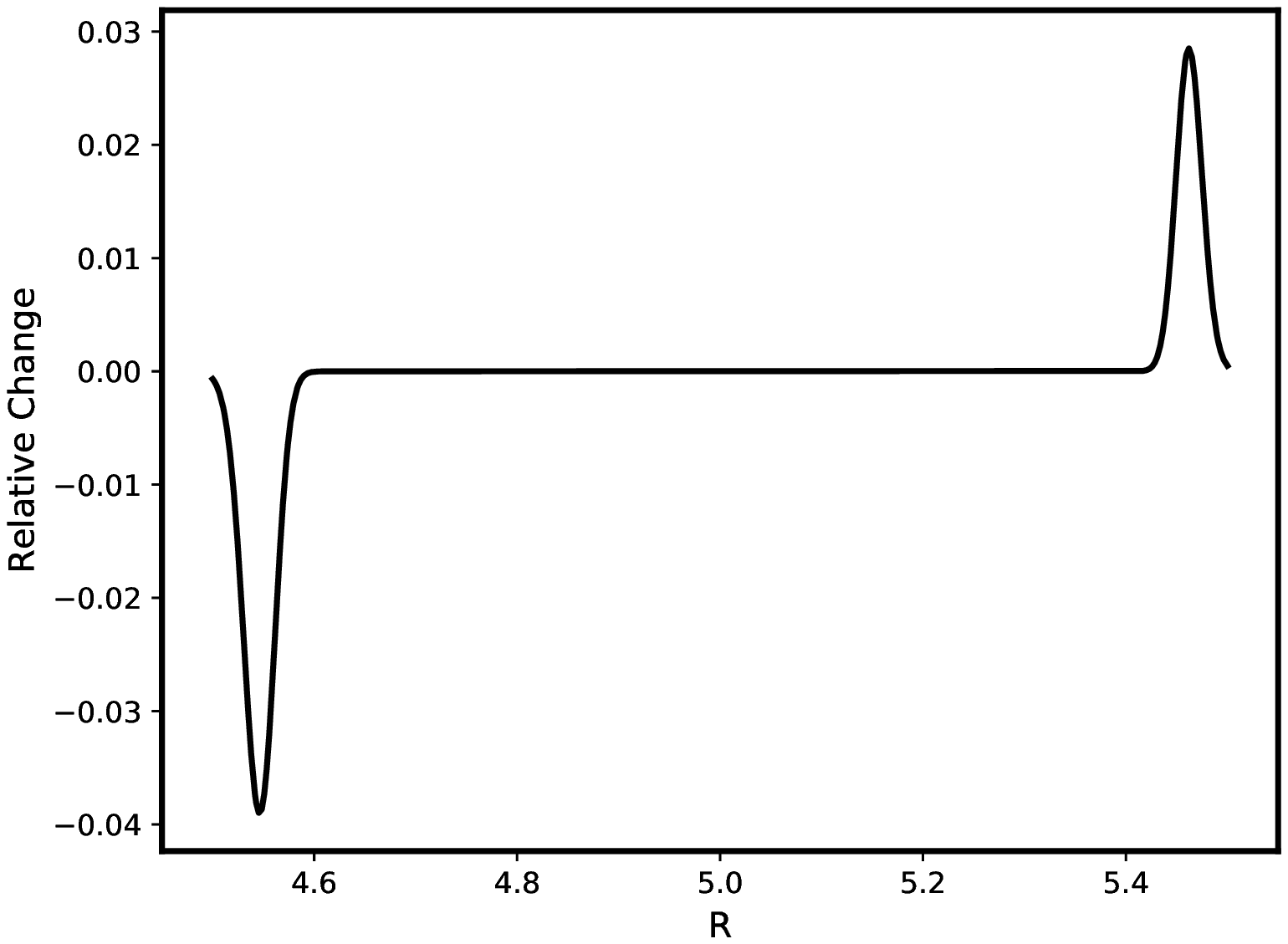}
  \label{fig:separate_psi_prof_effects_prof_2}
}

\subfigure[]{
  \includegraphics[width=0.35\textwidth, height=0.23\textheight]{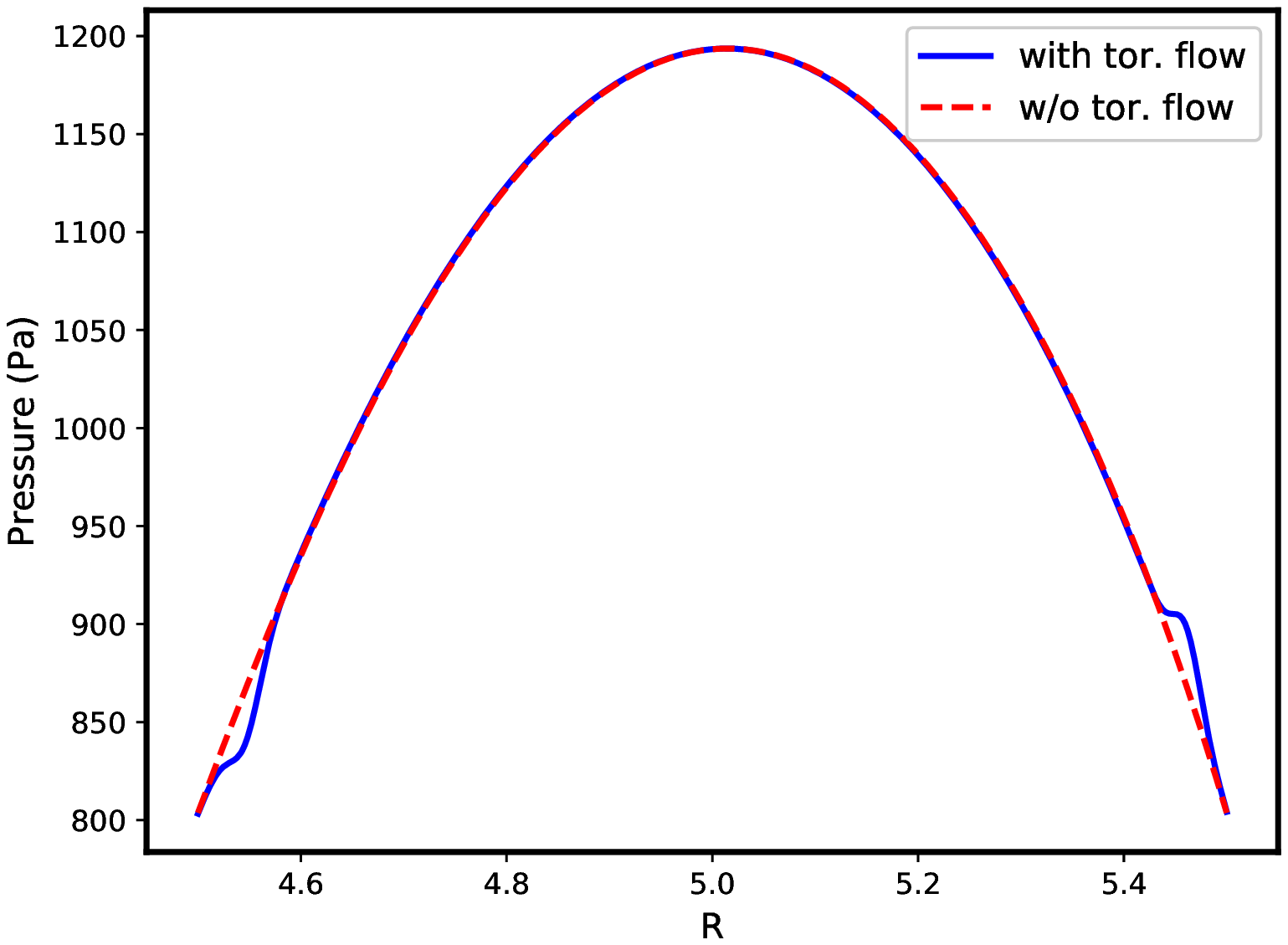}
  \label{fig:separate_psi_prof_effects_all_1}
}\subfigure[]{
  \includegraphics[width=0.35\textwidth, height=0.23\textheight]{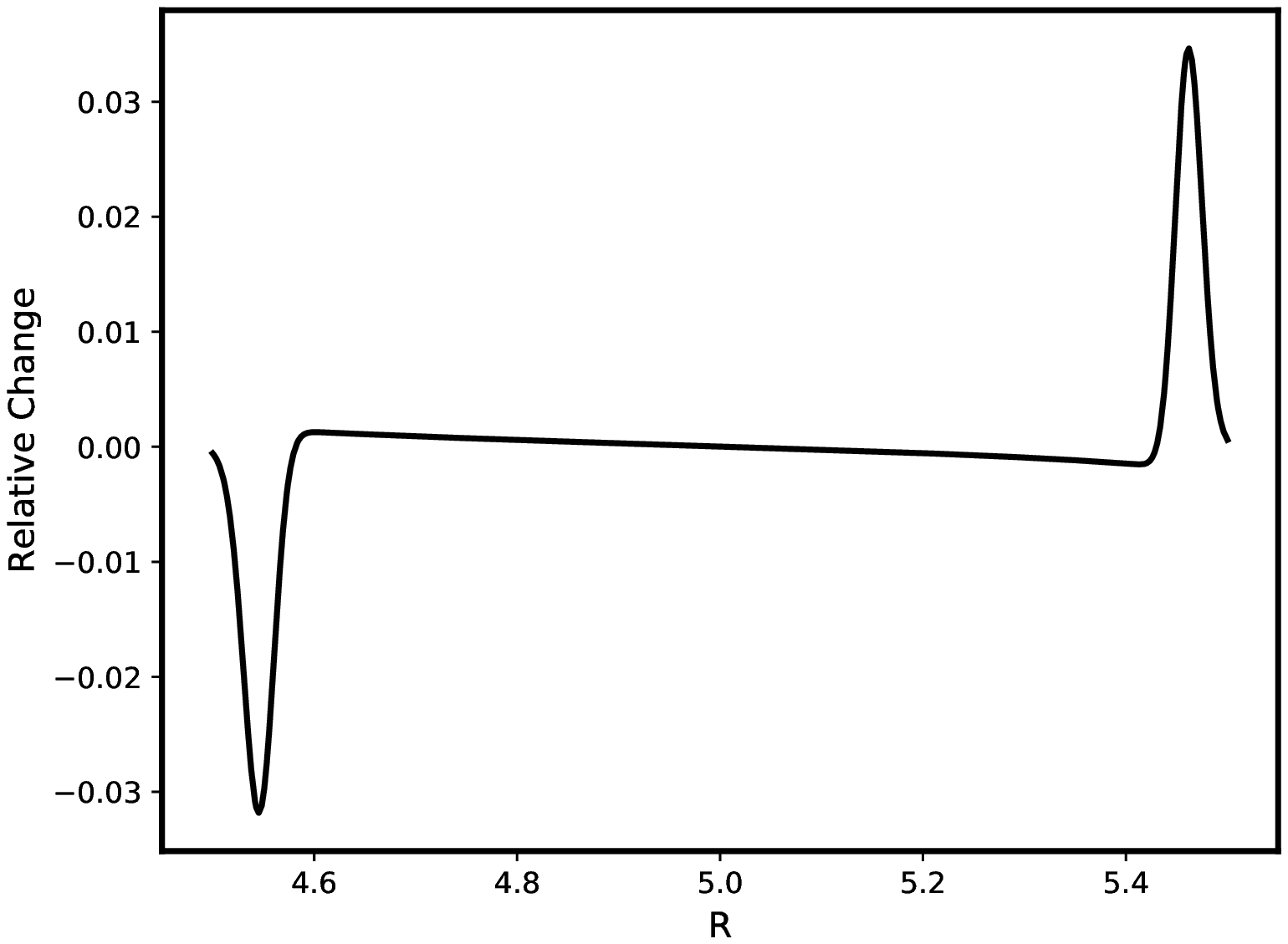}
  \label{fig:separate_psi_prof_effects_all_2}
}

\caption{(a) The normalized poloidal flux function $\psi$ from the NIMEQ solution for an L-model equilibrium in presence of a non-uniform toroidal flow with Gaussian profile, and (b) its change from the static equilibrium solution for different Mach numbers. The corresponding pressure and its relative change from the static equilibrium as functions of the major radius (c-d) ignoring and (e-f) including self-consistently the change in $\psi$ due to toroidal rotation, with the toroidal flow profile parameters $M=0.4$ and $\sigma=0.05$.}

\label{fig:separate_psi_prof_effects}
\end{figure}

When the toroidal rotation is localized to the edge region, as is often observed in H-mode experiments, the outward shift of equilibrium pressure profile becomes more localized in the same edge region as well, leading to the formation of a pedestal-like structure in the edge pressure profile. To demonstrate this, a toroidal flow with the Gaussian profile $\Omega = \Omega_{0} \exp[-(\psi-\mu)^2/2\sigma^2]$ is considered, where $\Omega_{0}$ denotes the peak amplitude of toroidal flow, $\mu$ the peak position, and $\sigma$ the width of the flow profile. For a low $\beta$ equilibrium, the change of $\psi$ induced by toroidal flow may be ignored according to Eq. (\ref{eq:terms_compare_gs}), which is also confirmed in Figs. \ref{fig:scan_M0_analy_myself}(a)-\ref{fig:scan_M0_analy_myself}(b) for a uniform flow and later in Figs. \ref{fig:separate_psi_prof_effects_psi_1}-\ref{fig:separate_psi_prof_effects_psi_2} for the nonuniform flow with Gaussian profile. Thus, the gradient of pressure may be written as
\begin{eqnarray}
\nabla P = \left[ \nabla P_{0} + 2 \frac{\mu - \psi}{\sigma^{2}} M^{2} \left( \frac{R^{2}}{R_{0}^{2}} - 1 \right) \nabla \psi +\frac{2R}{R_{0}^{2}} \hat{e}_{R} \right] \exp \left[ M^{2} \left( \frac{R^{2}}{R_{0}^{2}} - 1 \right) \right]
\label{eq:grad_P}
\end{eqnarray}
For a finite Mach number, the gradient of pressure approaches infinity as the width of the flow profile narrows down to zero, i.e. $|\nabla P| \rightarrow \infty$ when $\sigma \rightarrow 0$.

To further elaborate on this rotational effect on equilibrium, we consider a static L-mode equilibrium with pressure profile $P_{0}(\psi)=P_{axis}(1-\psi^{2})^{2}$ and safety profile $q=q_{axis}(1+(q_{edge}/q_{axis} -1)\psi^{4})$. The equilibrium solver NIMEQ is used for the numerical solution, which is recently extended to self-consistently take into account of the effects of toroidal flow with arbitrary profile \cite{nimeq_my_2020}. The computational domain is a circular grid with major radius $R_{0}=5.0m$ and minor radius $a=0.5m$.

\begin{figure}[ht]

\subfigure[]{
  \includegraphics[width=0.35\textwidth, height=0.23\textheight]{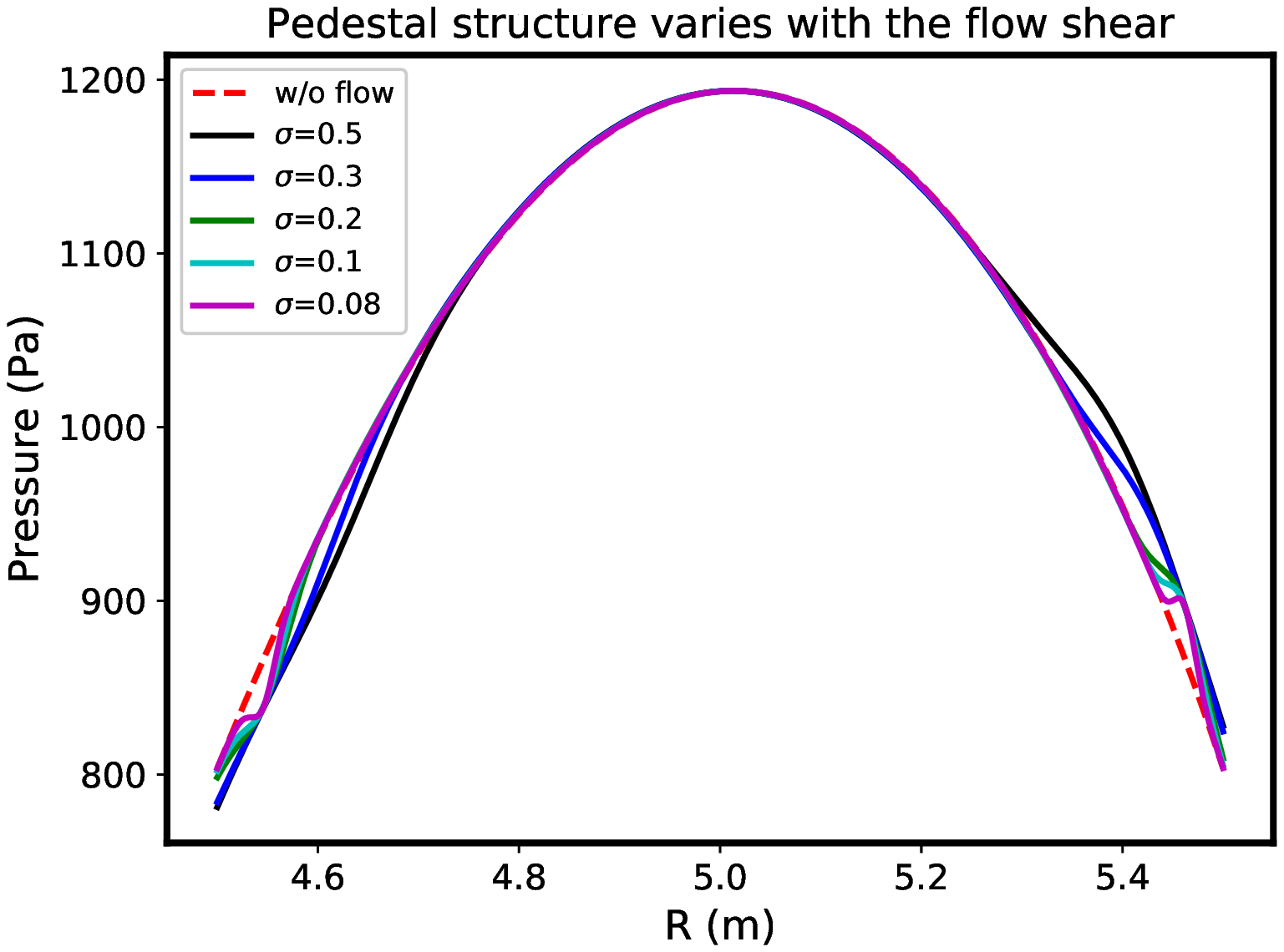}
  \label{fig:scan_theta_all}
}\subfigure[]{
  \includegraphics[width=0.35\textwidth, height=0.23\textheight]{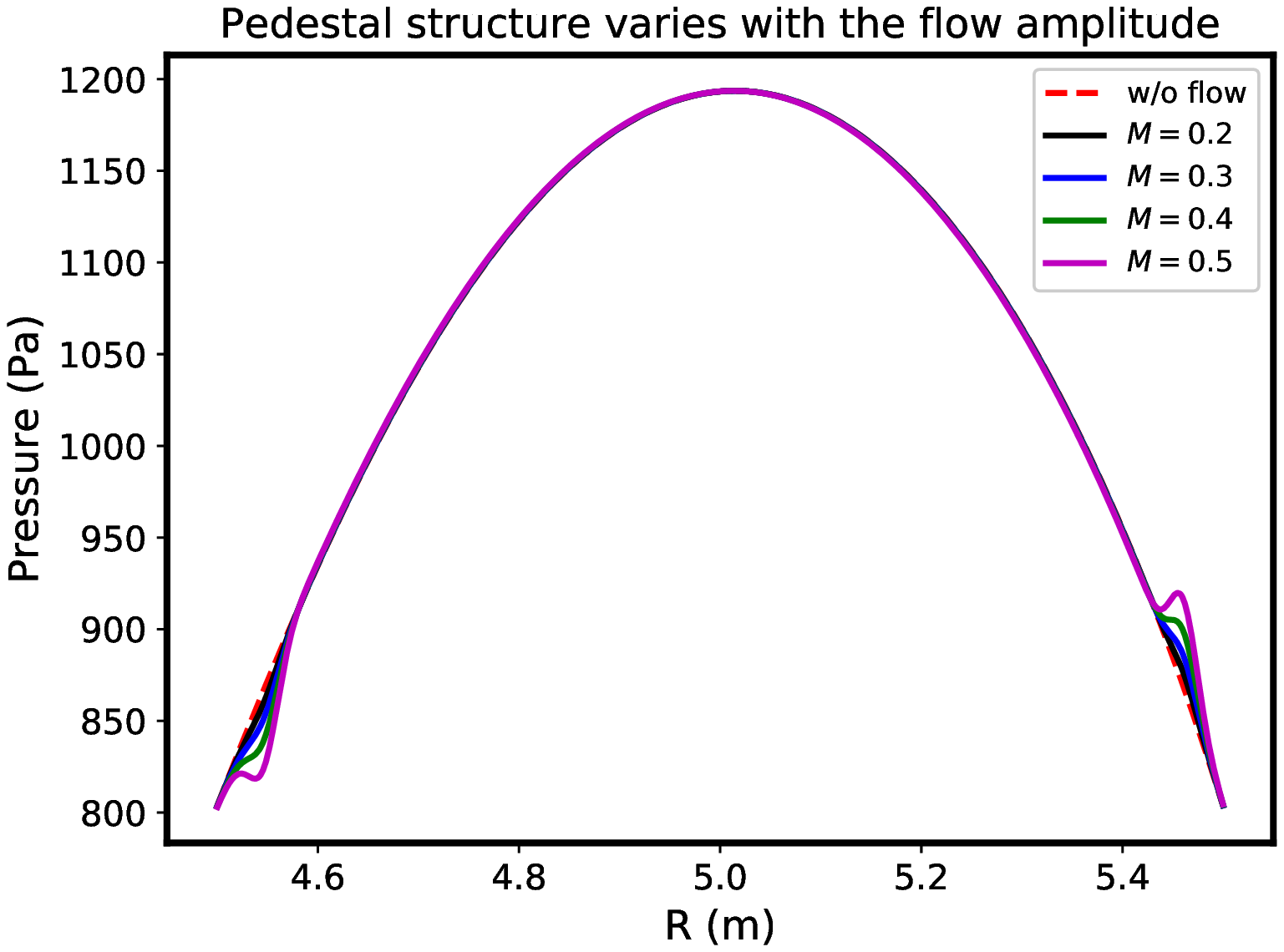}
  \label{fig:scan_amp_all}
}

\subfigure[]{
  \includegraphics[width=0.35\textwidth, height=0.23\textheight]{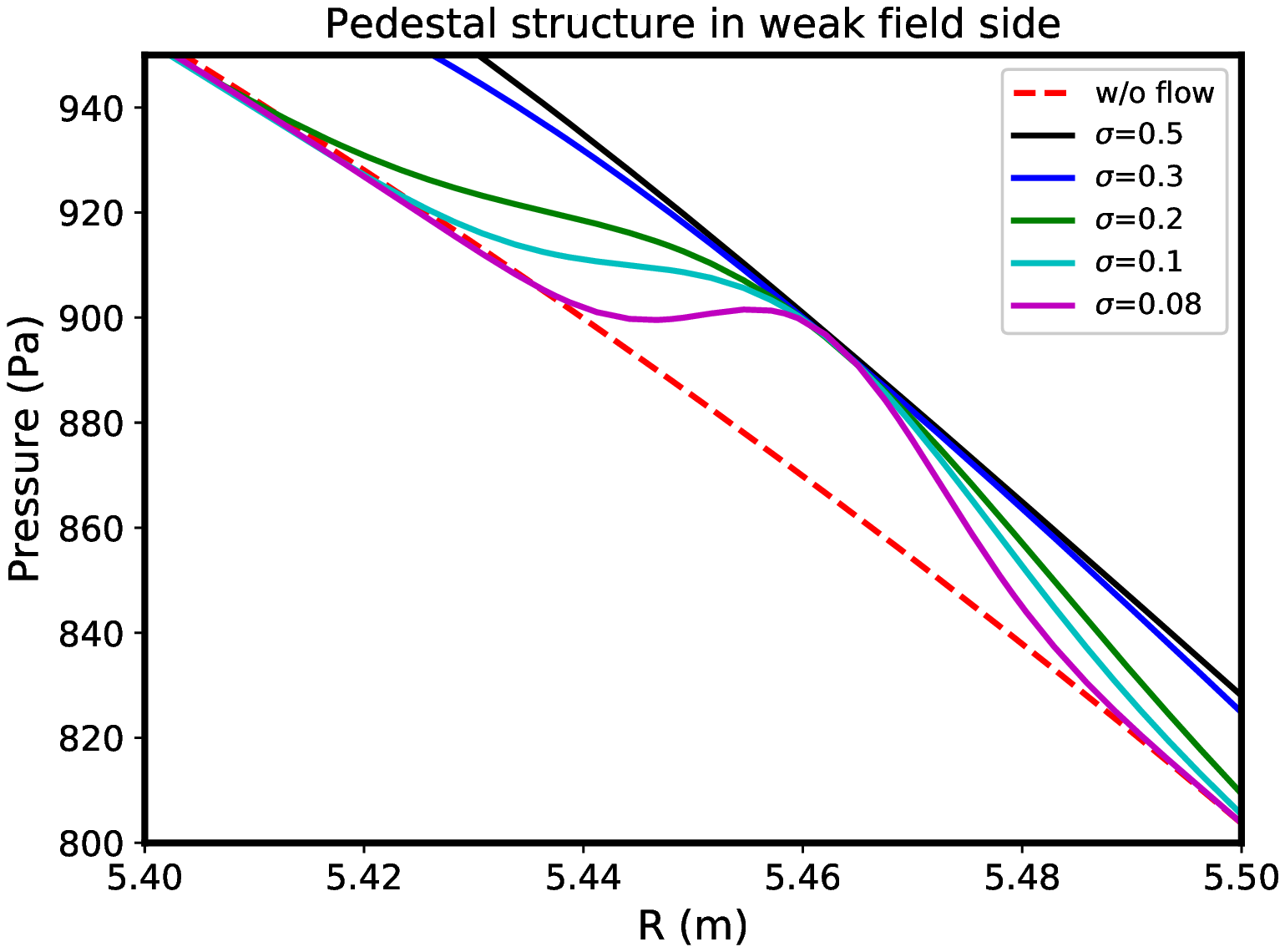}
  \label{fig:scan_theta_RHS}
}\subfigure[]{
  \includegraphics[width=0.35\textwidth, height=0.23\textheight]{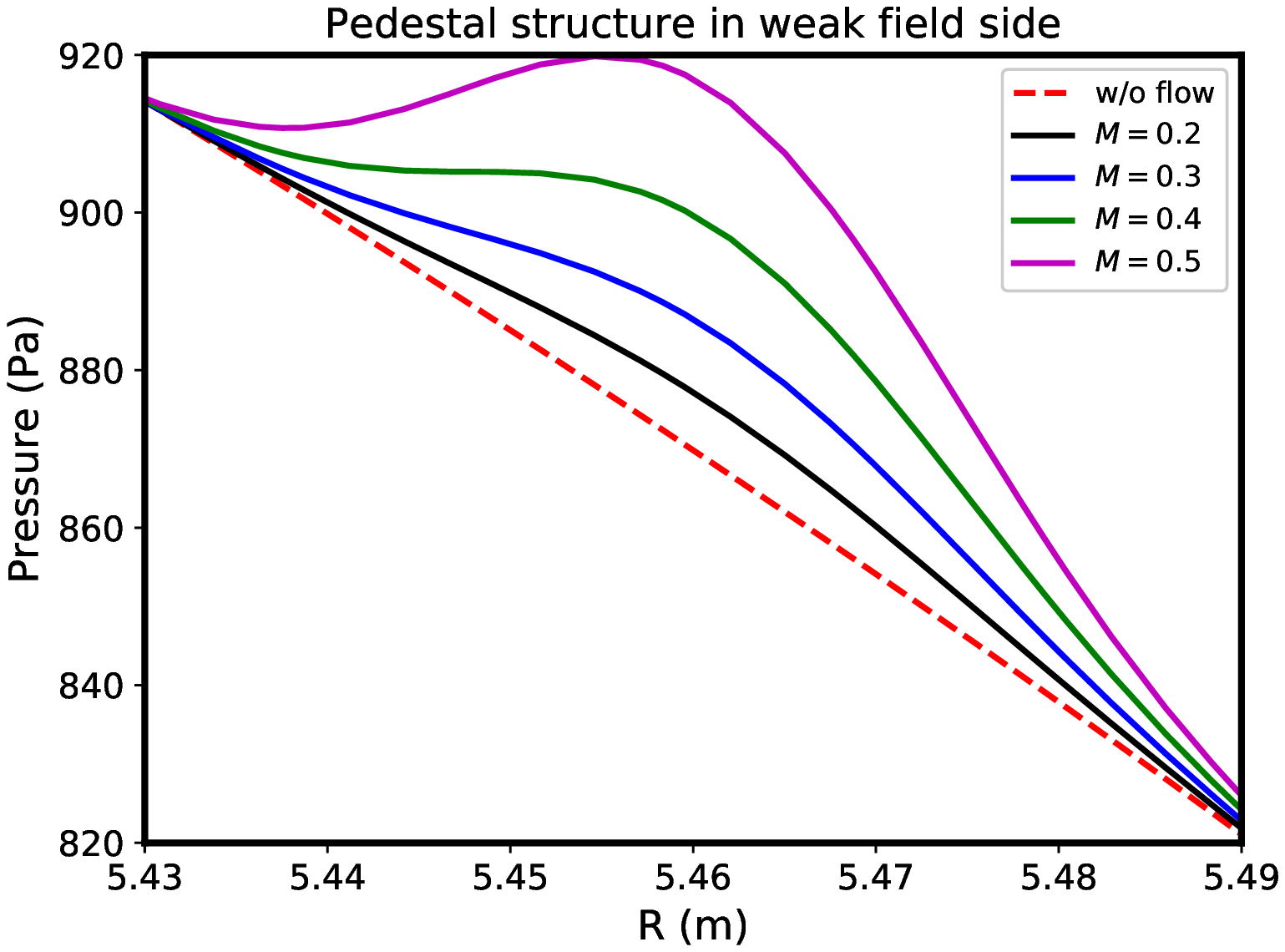}
  \label{fig:scan_amp_RHS}
}

\subfigure[]{
  \includegraphics[width=0.35\textwidth, height=0.23\textheight]{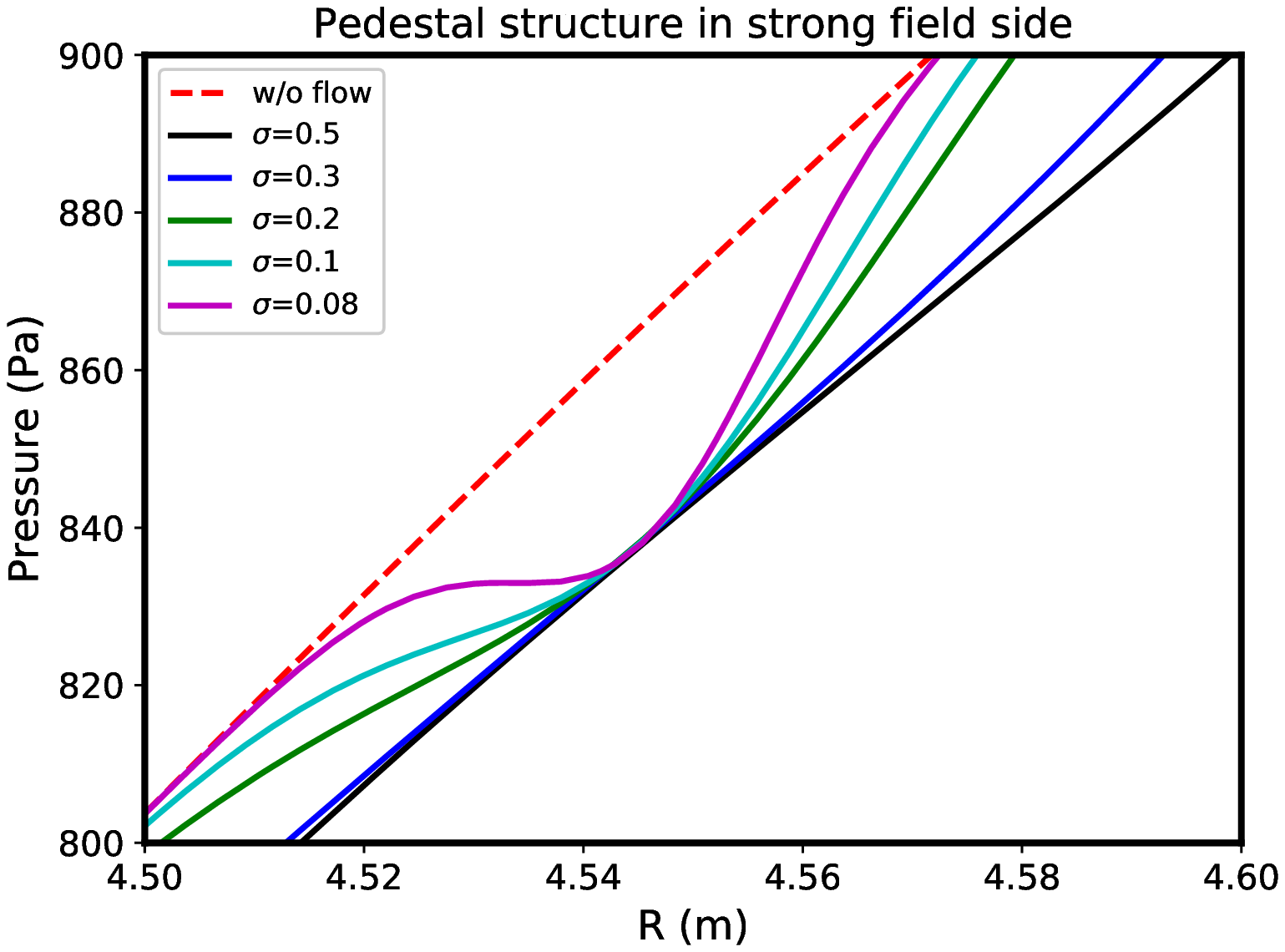}
  \label{fig:scan_theta_LHS}
}\subfigure[]{
  \includegraphics[width=0.35\textwidth, height=0.23\textheight]{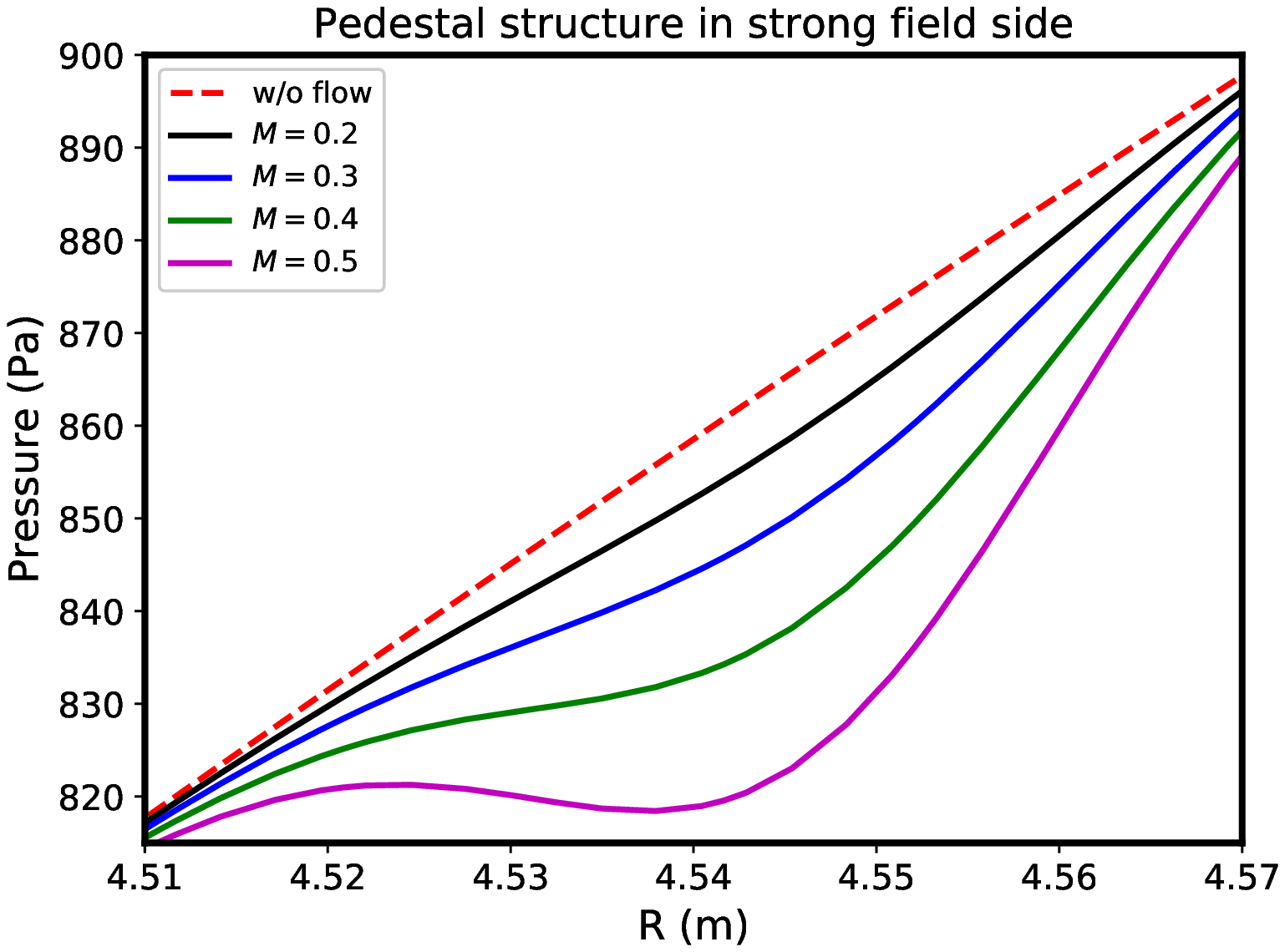}
  \label{fig:scan_amp_LHS}
}

\caption{The pressure from the NIMEQ solution as functions of the major radius for Gaussian flow profiles with different (a) profile width $\sigma$, and (b) peak Mach number $M$. The weak field and strong field sides of (a)-(b) are zoomed in (c)-(d) and (e)-(f) respectively.}

\label{fig:scan_pedestal}
\end{figure}

\begin{figure}[ht]

\subfigure[]{
  \includegraphics[width=0.5\textwidth]{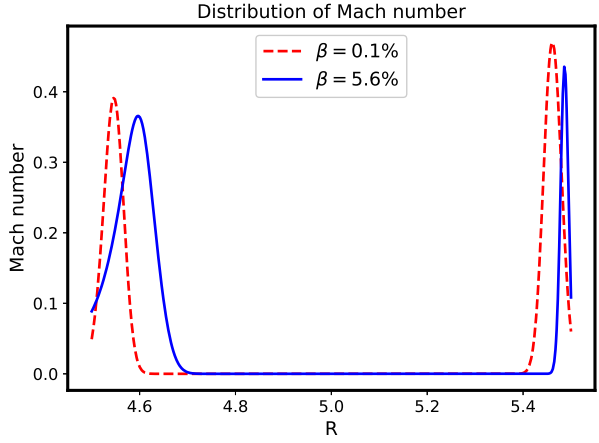}
  \label{fig:compare_mach_low_high}
}\subfigure[]{
  \includegraphics[width=0.5\textwidth]{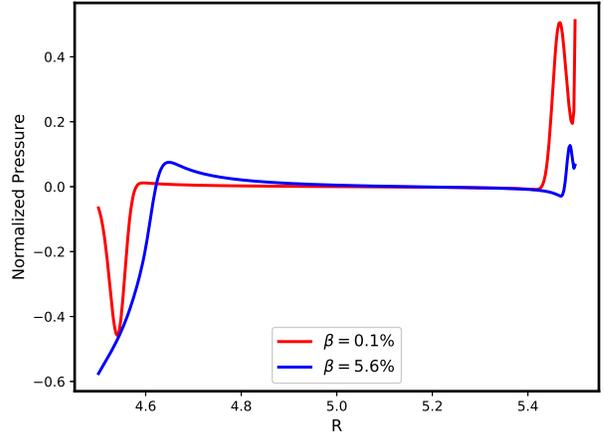}
  \label{fig:scan_beta_pedestal_ratio}
}

\caption{(a) The Mach number and (b) the relative change of pressure due to the presence of a non-uniform toroidal rotation as a function of the major radius in a lower $\beta$ (0.1\%) and higher $\beta$ (5.6\%) equilibrium.}
\label{fig:scan_beta_pedestal}
\end{figure}

A Gaussian flow profile with $M=0.4$, $\mu=0.6$, and $\sigma=0.05$ is considered first. If the equilibrium magnetic flux $\psi$ is kept same as the static equilibrium, the Gaussian flow profile alone, upon substitution into Eq. (\ref{eq:nimeq_p}), is able to introduce a pedestal-like structure in the pressure profile near the plasma edge (Figs. \ref{fig:separate_psi_prof_effects_prof_1}-\ref{fig:separate_psi_prof_effects_prof_2}). When the equilibrium flux $\psi$ is numerically solved self-consistently including the toroidal flow with Gaussian profile, its change from the static equilibrium is quite small, similar to the uniform toroidal flow case (Figs. \ref{fig:separate_psi_prof_effects_psi_1}-\ref{fig:separate_psi_prof_effects_psi_2}). As a result, the pedestal-like structure in pressure profile remains intact when the effects of toroidal flow with Gaussian profile are fully taken into account through the self-consistent equilibrium solution (Figs. \ref{fig:separate_psi_prof_effects_all_1}-\ref{fig:separate_psi_prof_effects_all_2}). Further variations of the equilibrium solutions over a range of parameters $\sigma$ and $M$, show that the edge pedestal-like structure in pressure profile becomes more pronounced as the peaking and the localization of flow profile enhance (Fig. \ref{fig:scan_pedestal}). For higher $\beta$ equilibrium, the larger Shafranov shift tends to squeeze both the Gaussian flow profile and the pedestal-like structure in pressure profile on the weak field side in the major radius direction (Fig. \ref{fig:scan_beta_pedestal}).

\section{$q$-profile flattening and reversal of local magnetic shear due to toroidal flow}
\label{sec:q_flatted}

For a given $F(\psi)$, the presence of toroidal flow in a tokamak equilibrium can alter the profile of safety factor $q$ through its effects on the poloidal magnetic flux, since
\begin{eqnarray}
q(\psi,M_{0})=\frac{\mu_{0}F(\psi)}{2\pi} \oint_{\psi} \frac{1}{|\nabla \psi|(R,Z,M_{0})} \frac{\mathrm{d}l}{R}=\frac{\mu_{0}F(\psi)}{2\pi} f(\psi,M_{0})
\label{eq:q_F_omg}
\end{eqnarray}
where the line integral is along the poloidal flux surface \cite{freidberg_ideal_2014}. Previous study finds that the off-axis flow can change $q$ profile and generate the negative magnetic shear \cite{lee2019}. Further insight on the dependence of $q$ profile on the toroidal flow may be gained from the relation.

\begin{eqnarray}
q(\psi) = \frac{F(\psi)^{2}}{4\pi^{2} \mu_{0} \langle J_{\parallel} \rangle} \left[ \frac{V'(\psi)}{F(\psi)} \left \langle \frac{|\nabla \psi|^{2}}{R^{2}} \right \rangle \right]'
\label{eq:q_jpar}
\end{eqnarray}
where the flux surface average $\langle A \rangle=\frac{\int A \mathrm{d} l}{\int \mathrm{d}l}$ is defined as the integral over a constant $\psi$ surface, $A'=\frac{d A}{d \psi}$, and $V'(\psi)=2\pi \oint_{\psi} \frac{R\mathrm{d}l}{|\nabla \psi|}$ \cite{2010jardin}. Since the safety factor $q$ profile is inversely proportional to the parallel current density $\langle J_{\parallel} \rangle $, a non-monotonic $q$ profile may form when $J_{\parallel}$ is peaked off-axis.

When the toroidal flow is taken into account in the equilibrium self-consistently, the parallel current profile of rotational tokamak equilibrium can be derived as
\begin{eqnarray}
\langle J_{\parallel} \rangle = \frac{\mathrm{d}F}{\mathrm{d}\psi} \langle B \rangle + F \left \langle \frac{|\nabla P + \rho \Omega^{2} R \hat{e}_{R}|}{RBB_{p}} \right \rangle
\label{eq:jpar_M_w}
\end{eqnarray}
where $B_{p}$ is the poloidal magnetic field strength. Thus the toroidal flow can affect the parallel current, in a way similar to the pressure gradient. Since the change of $\psi$ induced by toroidal flow is negligible, as indeed indicated from the analysis and calculations in Sec.\ref{sec:pedestal_formation}, as a result, a localized and peaked toroidal flow profile may induce non-monotonic structures in the parallel current density as well as the safety factor profiles, as shown in the numerical examples below.

\begin{figure}[ht]
  \begin{center}
  \includegraphics[width=1\textwidth,height=0.6\textheight]{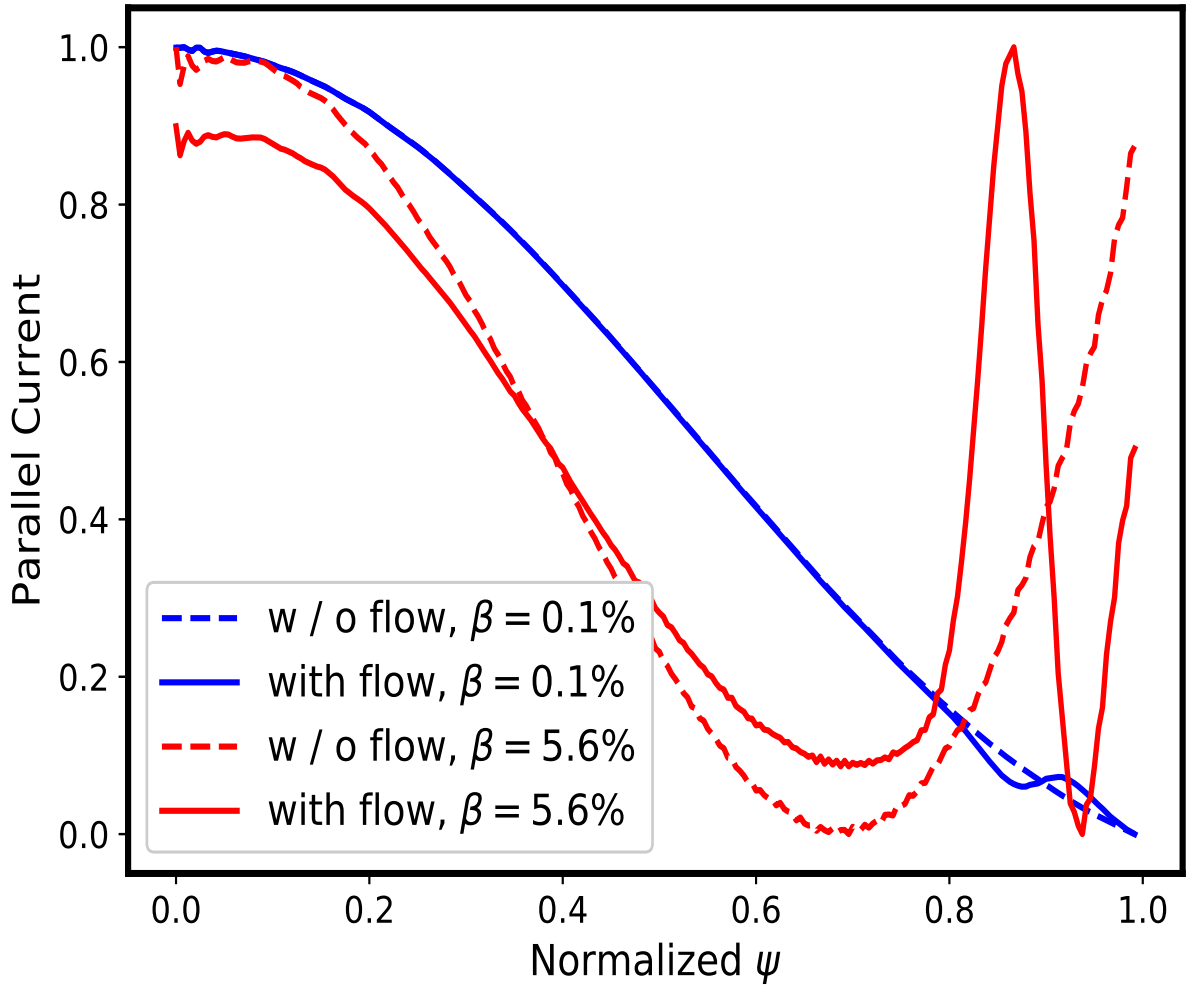}
  \end{center}

\caption{The parallel current density as a function of the normalized poloidal flux function $\psi$ in a lower $\beta$ (0.1\%, color blue) and a higher $\beta$ (5.6\%, color red) equilibrium in the presence (solid line) and the absence (broken line) of a non-uniform toroidal flow ($M=0.4$, $\sigma=0.05$).}
\label{fig:scan_beta_q_flatted_reversed_jpar}
\end{figure}

\begin{figure}[ht]

\subfigure[]{
  \includegraphics[width=0.7\textwidth]{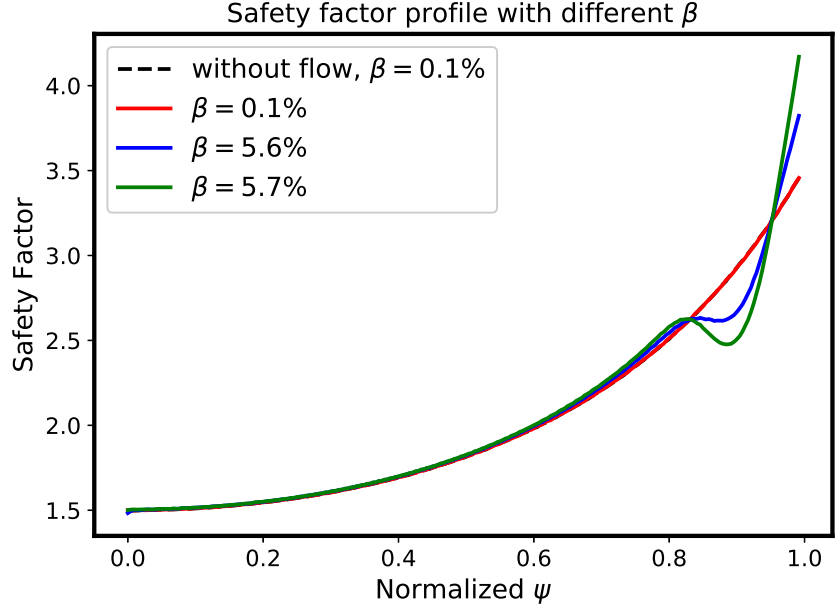}
  \label{fig:scan_beta_q_flatted_reversed}
}

\subfigure[]{
  \includegraphics[width=0.7\textwidth]{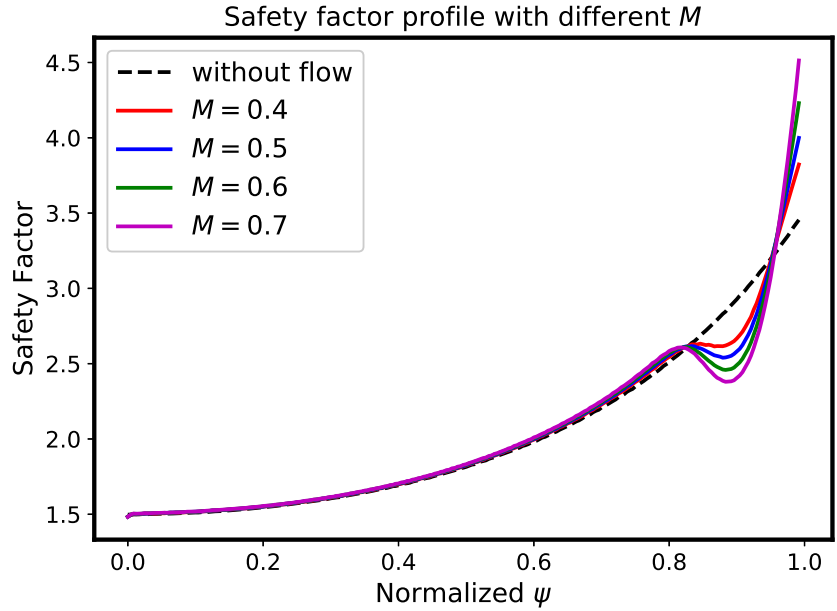}
  \label{fig:scan_mach_q_reversed}
}

\caption{The safety factor as a function of the normalized poloidal flux function $\psi$ in the absence (dark broken line) and the presence (solid lines) of a non-uniform toroidal flow with (a) a fixed peak Mach number $M=0.4$ and different core plasma $\beta$; (b) a higher $\beta$ (5.6\%) with different peak Mach numbers. $\sigma=0.05$ for all flow profiles.}

\end{figure}

For the same tokamak equilibrium and Gaussian flow profile considered in Sec.\ref{sec:pedestal_formation}, at lower $\beta=0.1\%$, there is a small bump in the parallel current profile at the peak location of the localized Gaussian toroidal flow. At higher $\beta=5.6\%$, the off-axis peak in the parallel current profile is dramatically enhanced due to the presence of the same toroidal flow profile and amplitude (Fig. \ref{fig:scan_beta_q_flatted_reversed_jpar}). This peak in parallel current profile reduces the safety factor around the peak position, leading to a local flattening of the $q$ profile and eventually a reversal of the magnetic shear as plasma $\beta$ further increases (Fig. \ref{fig:scan_beta_q_flatted_reversed}). In another word, the effects of toroidal flow on the parallel current density and $q$ profiles are amplified in higher $\beta$ regimes.

For a fixed $\beta$ or pressure profile, increasing the peak amplitude of toroidal rotation directly enhances the peak value of parallel current and the consequential reversal in magnetic shear (Fig. \ref{fig:scan_mach_q_reversed}). However, the required Mach number has to be much larger and even supersonic for the $q$-flattening or shear reversal regions to become substantial in the lower $\beta$ equilibrium.

\section{Summary and Discussion}
\label{sec:summary}

In summary, our analytical and numerical calculations find that the presence of a localized edge toroidal rotation can directly lead to the formation of a pedestal-like structure in pressure profile as an immediate consequence of the self-consistent tokamak equilibrium solution of the extended Grad-Shafranov equation including effect of toroidal flow. In general toroidal flow shifts plasma profile and flux surface outward in the direction of tokamak major radius. However, a purely rigid toroidal flow alone can not generate the edge pedestal-like structure. The enhanced local pressure gradient is shown to be inversely proportional to the width of a localized and peaked toroidal flow profile. In addition, such an edge localized toroidal rotation is able to induce an edge localized and peaked structure in the profile of the surface-averaged parallel current density, which in turn leads to the flattening of $q$-profile and even reversal of the local magnetic shear in the edge pedestal-like region, especially at higher $\beta$ regimes.

Thus, the formations of pedestal structure and weakened or reversed magnetic shear in the edge region, often associated with H- and QH-modes, are likely also a natural and self-consistent MHD tokamak equilibrium response to the presence of an edge localized toroidal rotation. Extension of this study to the more realistic non-circular shaped, divertor tokamak configurations may be necessary in order to bring the calculations here closer to experiment comparisons, which is planned for future work.

\section*{Acknowledgments}
This work was supported by the Fundamental Research Funds for the Central Universities at Huazhong University of Science and Technology Grant No. 2019kfyXJJS193, the National Natural Science Foundation of China Grant Nos. 11775221 and 51821005, and U.S. Department of Energy Grant Nos. DE-FG02-86ER53218 and DE-SC0018001. This research used the computing resources from the Supercomputing Center of University of Science and Technology of China.

\section*{Data Availability Statement}
The data that support the findings of this study are available from the corresponding author upon reasonable request.

\newpage

\newpage

\newpage

\newpage
\newpage
\newpage

\newpage
\section{Reference}
\bibliographystyle{unsrt}

\end{document}